\documentclass[a4paper]{llncs}
\usepackage{times,cite}
\usepackage[T1]{fontenc}
\usepackage[utf8]{inputenc}
\usepackage{makeidx}
\usepackage{amssymb,amstext}
\usepackage{tikz,pgfplots}
\pgfplotsset{compat=newest}
\usetikzlibrary{positioning}
\usetikzlibrary{shapes.geometric,calc,matrix,arrows,decorations.pathmorphing}
\usetikzlibrary{3d,shapes,snakes}
\usetikzlibrary{shadings,mindmap,trees}
\tikzset{
	braces/.style = {
		outer sep=-1pt,
		left delimiter=(,
		right delimiter=),
		align=center,
	},
}

\newdimen\fwd

\newcommand{\R}{\mathbb{R}}
\newcommand{\SQN}{\ensuremath{\mathrm{S}q\mathrm{N}}}
\begin{document}
\newpage

\author{Kai~Brehmer\inst{1} \and
	Benjamin~Wacker\inst{1} \and
	Jan~Modersitzki\inst{1,2}}
\authorrunning{Kai~Brehmer et al.}
\tocauthor{Kai Brehmer, Benjamin Wacker, and Jan Moderistzki}
\title{Simultaneous Registration of Image Sequences -- 
	a novel singular value based images similarity measure}
\titlerunning{Registration of Image Sequences}
\institute{Institute of Mathematics and Image Computing, 
	University of L{\"u}beck, Germany, \\ \email{brehmer@mic.uni-luebeck.de} \and
	Fraunhofer MEVIS, L{\"u}beck, Germany}

\maketitle

\begin{abstract}
	The comparison of images is an important task in image processing. 
	For a comparison of two images, a variety of measures has been suggested.
	However, applications such as dynamic imaging or serial sectioning provide
	a series of many images to be compared. 
	When these images are to be registered, the standard approach is to sequentially 
	align the~$j$-th image with respect to its neighbours and sweep with respect to~$j$. 
	One of the disadvantages is that information is distributed only locally.
	
	We introduce an alternative so-called \SQN\ approach.
	\SQN\ is based on the Schatten-$q$-norm of the image sequence gradients,
	i.e. rank information of image gradients of the whole image sequence.
	With this approach, information is transported globally.
	Our experiments show that \SQN\ gives at least comparable registration results 
	to standard distance measures but its computation is about six times faster.
	
\end{abstract}

\section{Introduction}

Fusion and comparison of images is an important task in image processing.
Examples include motion correction in Dynamic Contrast Enhanced Magnetic 
Resonance Imaging (DCE-MRI) or reconstruction in Histological Serial Sectioning (HSS). 
These tasks require a suitable image similarity measure which takes application 
dependent image features into account. For example, in DCE-MRI or HSS
the measure has to be invariant under intensity variations due to 
contrast uptake or staining artefacts, respectively.

For pairs of two images, a variety of options has been proposed and is well-understood. 
Among the various choices are~$L_2$ based measures, normalized gradient fields (NGF), 
mutual information or Kullback-Leibler divergence; 
see, e.g.~\cite{modersitzki04,modersitzki09} and references therein.
For scenarios where more than two images~$I_t$, $t=1,\ldots,T$, are to be aligned, 
the standard approach is to adapt the pairwise procedure in a sequential fashion. 
More precisely, the~$t$-th image is aligned with respect to the neighbours 
$I_{t-1}$ and~$I_{t+1}$ for all~$t$. Since the correct aligned neighbours are 
yet unknown, the procedure has to be repeated until convergence.

This paper proposes an alternative so-called \SQN\ approach, which is based on
the rank of image gradients of the whole sequence. A key feature of this new
approach is that similarity information is made globally. The approach is
inspired by work of Möllenhoff~et.~al.~\cite{moellenhoff15} and Haber et.~ al.~\cite{haber05}.
In~\cite{moellenhoff15}, Schatten-$q$-Norms are used for color image denoising
and in~\cite{haber05} local normalized gradient fields are introduced as for pairwise image registration.
In our paper, \SQN\ is used as a data fitting term and globally
normalized image gradients of the whole sequence with arbitrary many images
are used as a starting point.

We derive and motivate the \SQN\ similarity measure.
We also compare the performance to state-of-the-art 
measures on real life data.
Our computations for a serial sectioning of a mouse brain indicate
that \SQN\ results are qualitatively comparable to a sequential 
$L_2$ based registration but can be obtained about six times faster;
see also~\cite{brehmer18}.


\section{The novel similarity measure SqN}

The basic idea of our novel measure derives from color image denoising~\cite{moellenhoff15}. 
There, the linear dependency of the gradients of the three color channels is used for regularization.
This dependency is quantified by a Schatten-$q$-Norm~\cite{bhatia13}.
We generalize this idea and use it in the context of image registration.
In particular, we extend the idea to an arbitrary number of images.
Moreover, we also use normalized gradient fields rather than image gradients.
Finally, we use the functional as a datafit rather than a regularizer.

The Schatten-$q$-Norm of a matrix is essentially the~$q$-norm of the 
vector of its singular values~\cite{bhatia13}.
More precisely, for any~$A \in \mathbb{R}^{n,T}$ there exists a 
singular value decomposition (SVD)~\cite{golub12}, 
\(
	A = U\Sigma V^{\top},
	\text{ with } U^{\top}U = E_n,\
	V^{\top}V = E_T,
\)
where~$E_d$ denotes the~$d$-by-$d$ identity matrix,
$\Sigma\in\mathbb{R}^{n,T}$ is a diagonal matrix with diagonal entries
$\sigma_1,\ldots,\sigma_{\min\{n,T\}}$ and 
$\sigma_j\ge\sigma_{j+1}\ge0$. 
The Schatten-$q$-(quasi)-norm of~$A$ is then defined as
\[
  \|A\|_{S,q}^{q} 
  :=\|\Sigma\|_{S,q}^{q} 
  :=
  \sum_i\sigma_{i}^{q}
  \text{ for }
  q \geq 0.
\]

We assume that~$d$ dimensional images~$I^t\in\mathbb{R}^n$ are given,
where~$t=1,\ldots,T$ and the spatial dimension is denoted by~$n=m_1 \times \dots \times m_d$.
Following~\cite{haber05}, regularized normalized gradients are given by
\[
	\eta^t:=\nabla I^t/\|\nabla I^t\|_\theta\in\R^{dn},
	\text{ where }
	\|\nabla I^t\|_\theta:=\sqrt{\|\nabla I^t\|_{\R^{dn}}^2+\theta^2}.
\]
Here, the parameter~$\theta$ discrimiantes signal from noise;
see~\cite{haber05} for a discussion of choices of~$\theta$ and discretization issues.
Setting~$I=(I^1,\ldots,I^T)$ we define the new image similarity measure:
\[
	\SQN:\R^{n,T}\to\R,\quad
	\SQN(I):=
	\left\|\Big[\
		\frac{\nabla I^1}{\|\nabla I^1\|_\theta},\ldots,
		\frac{\nabla I^T}{\|\nabla I^T\|_\theta}
		\ \Big]\right\|_{S,q}^{q}.
\]
The registration model is to minimize
$J(y):=\SQN(I \circ y)+\sum_{t=1}^T S(y^t)$,
with~$y=(y^1,\ldots,y^T)$ and~$y^t:\R^d\to\R^d$; cf.~\cite{modersitzki09}.


\begin{figure}[htbp]\centering\fwd=17mm
	\begin{tabular}{cccc}
		& axial & coronal & sagittal
		\\[0mm]
		\rotatebox{90}{pre-aligned}
		&\includegraphics[height=\fwd,angle=90,origin=center]%
		{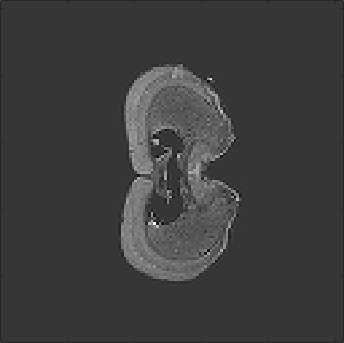}
		&\includegraphics[height=\fwd]%
		{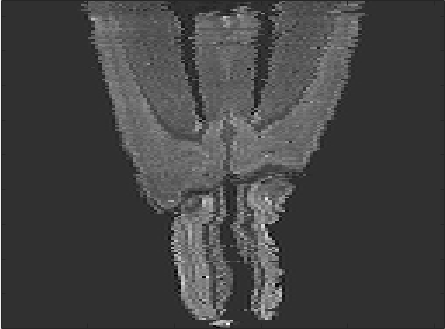}
		&\includegraphics[width=\fwd,angle=90]%
		{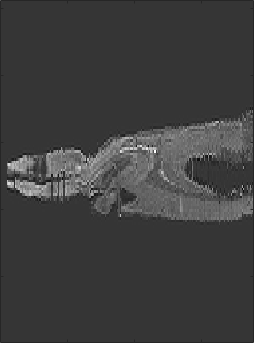}
		\\[0mm]
		\rotatebox{90}{\hskip4.5mm \SQN}
		&\includegraphics[height=\fwd,angle=90,origin=center]%
		{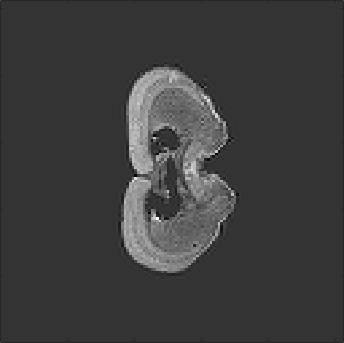}
		&\includegraphics[height=\fwd]%
		{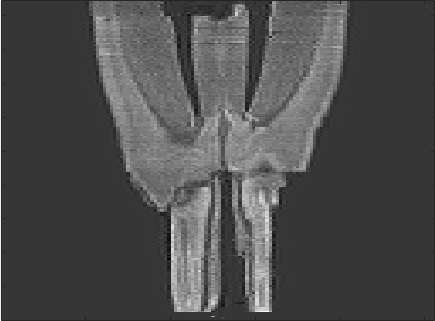}
		&\includegraphics[width=\fwd,angle=90]%
		{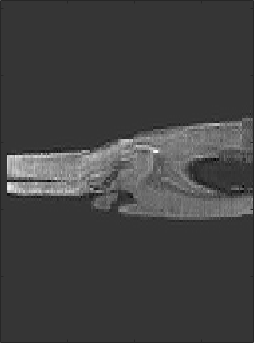}
		\\[0mm]
		\rotatebox{90}{\hskip4.5mm $L_2$}
		&\includegraphics[height=\fwd,angle=90,origin=center]%
		{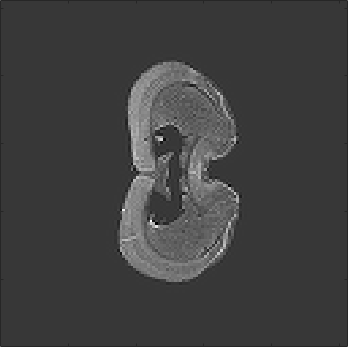}
		&\includegraphics[height=\fwd]%
		{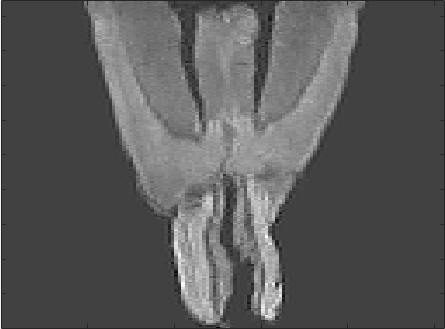}
		&\includegraphics[width=\fwd,angle=90]%
		{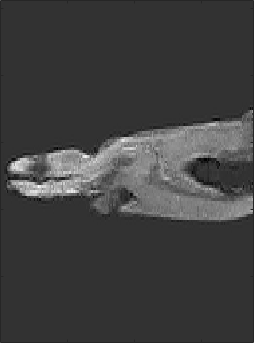}
		\\[0mm]
	\end{tabular}
	
	\caption{%
		Registration results for a histological serial sectioning of 
		mouse brain;
		data courtesy of O.~Schmitt, University of Rostock, Germany.
		Representative axial, coronal, and sagittal slices of 
		the 3D data of size 256-by-256-by-189 are shown.
		Displayed are linearly pre-aligned data (top row),
		\SQN-registered data (middle row), and 
		$L_2$-registered data (bottom row).
	}
	\label{fig:results_HNSP}
\end{figure}

\section{Numerical Results and Discussion}

We present results for a histological serial sectioning of a sectioned mouse brain, data courtesy of Oliver Schmitt. For results on DCE-MRI data we refer to~\cite{brehmer18}.
Fig.~\ref{fig:results_HNSP} shows results for a sequential linear
pre-registration, the new \SQN\ based registration, and a standard sequential
$L_2$ based registration as a reference (robust and fast to compute).

Although only one iteration was performed for the sequential~$L_2$ registration, the computing time is about six times as for the \SQN\ approach. The \SQN\ result also shows a much stronger spatial correlation, 
indicating that the sequential approach has not yet converged.

Our future work addresses the optimal choice of~$q$ 
(currently~$q=0.5$ as in~\cite{moellenhoff15}) and the extension to 3D.

\paragraph{Acknowledgement}
The authors acknowledge the financial support by the Federal Ministry of Edu\-cation and Research of Germany in the framework of MED4D (project number 05M16FLA)


\end{document}